# The Future of Sustainability in Germany: Areas for Improvement and Innovation

**Mehrnaz Kouhihabibi**

Department of Letter and Philosophy, Sapienza University, Italy.

**Erfan Mohammadi**

Faculty of Entrepreneurship, University of Tehran, Tehran, Iran

**Abstract:** This paper reviews the literature on biodegradable waste management in Germany, a multifaceted endeavor that reflects its commitment to sustainability and environmental responsibility. It examines the processes and benefits of separate collection, recycling, and eco-friendly utilization of biodegradable materials, which produce valuable byproducts such as compost, digestate, and biogas. These byproducts serve as organic fertilizers, peat substitutes, and renewable energy sources, contributing to ecological preservation and economic prudence. The paper also discusses the global implications of biodegradable waste management, such as preventing methane emissions from landfills, a major source of greenhouse gas, and reusing organic matter and essential nutrients. Moreover, the paper explores how biodegradable waste management reduces waste volume, facilitates waste sorting and incineration, and sets a global example for addressing climate change and working toward a sustainable future. The paper highlights the importance of a comprehensive and holistic approach to sustainability that encompasses waste management, renewable energy, transportation, agriculture, waste reduction, and urban development.

*Keywords*: Sustainability, Biodegradable waste management, Environmental responsibility, Recycling Climate protection.

## 1. Introduction

Sustainability has emerged as a paramount concern gaining increasing significance in recent years as the nation endeavors to curtail its environmental impact and champion sustainable practices (Abbas & Sağsan, 2019 Jahanshahi et al., 2020; Hessari et al., 2022; 2022; Mohammadi et al., 2023). Renowned for its longstanding environmental policy leadership, Germany's sustainability advocacy dates back to the 1970s (Abdelkafi & Täuscher, 2016). The contemporary emphasis on sustainability has intensified as Germany confronts the formidable challenges posed by climate change, while fervently aspiring to forge a sustainable future for its citizens (Al-Salem & Lettieri, 2021). Evidencing its dedication to sustainability, are the myriad policies and initiatives meticulously devised and executed to foster sustainable living. Ranging from the profound shift towards renewable energy to innovative approaches in sustainable transportation and waste management (Moezzi et al., 2012; Dehkordy et la., 2013; Shamsaddini et al., 2015; Hessari & Nategh, 2022; Rezaei et al., 2022; Rouhani & Mohammadi, 2022), Germany has firmly established itself as a vanguard of eco-consciousness (Amlinger et al., 2021). Central to Germany's fervent pursuit of sustainability are various compelling drivers, encompassing a resolute commitment to reduce greenhouse gas emissions, advance energy efficiency, and zealously safeguard the environment. Of notable significance is Germany's epochal transition to renewable energy sources. The nation has set ambitious objectives, seeking to derive 80% of its electricity from renewable sources by 2050, thereby minimizing reliance on fossil fuels (Boldrin et al., 2009). This momentous shift has been buttressed by a constellation of visionary policies and initiatives, including the incentivizing feed-in tariffs that kindle the development of renewable energy sources and the Renewable Energy Act, which accords renewable energy producers a fixed feed-in tariff for two decades. Equally deserving of attention is Germany's concerted focus on sustainable transportation. This entails implementing a comprehensive array of policies and initiatives, embracing subsidies for electric and hybrid vehicles, expanding public transportation systems, and creating an extensive cycling infrastructure network (De Jaeger et al., 2011). Collectively, these endeavors have diminished the nation's dependence on fossil fuels while proffering sustainable transportation alternatives. Moreover, sustainability finds resonance in diverse sectors, encompassing agriculture, waste management, and urban development. Germany has forged a path toward sustainable agricultural practices, ardently supporting organic farming and concurrently undertaking measures to curtail waste and galvanize recycling (Jahanshahi et al., 2019; Khaksar et al., 2010; Asadollahi et al., 2011; Hessari et al., 2023). In urban areas, the nation has exhibited unwavering commitment to sustainable urban development, prioritizing public transportation networks, pedestrian-friendly infrastructure, and verdant communal spaces. Sustainability has transcended its status as an important issue to become an imperative concern in Germany, emerging as a countermeasure to the multifaceted challenges posed by climate change. This profound commitment is irrefutably exemplified by the panoply of policies and initiatives that have been strategically instituted to promulgate sustainable practices, encompassing a seismic shift to renewable energy, the evolution of sustainable transportation, and innovative waste management strategies. As the global community grapples with the inexorable ramifications of climate change, Germany's steadfast devotion to sustainability stands as a paragon for nations endeavoring to craft a more sustainable future. This paper reviews the literature on the various policies and initiatives that Germany has implemented to foster sustainable living in different domains, such as renewable energy, transportation, waste management, agriculture, and urban development. The paper also discusses the drivers and motivations behind Germany's commitment to sustainability, as well as the benefits and challenges of its approach. The paper aims to provide a comprehensive and holistic overview of Germany's sustainability efforts and to highlight its role as a global example for addressing climate change and working toward a sustainable future.

## 2. History of sustainability in Germany

2.1 Evolution of Sustainability in Germany

Germany's rich legacy of environmentalism and sustainable living finds its roots in the environmental activism of the 1970s. This pivotal era witnessed a burgeoning wave of citizen concern, which stemmed from mounting apprehensions over pollution, deforestation, and other pressing environmental issues (Hahn & Weidtmann, 2020). The collective activism of the populace during this time yielded the inception of various environmental organizations (Vesal et al., 2013; Sepahvand et al., 2023), including the noteworthy Greenpeace Germany, founded in 1980. As the years unfurled, Germany's steadfast dedication to environmental causes endured and deepened, with the early 1990s marking a notable turning point as the concept of sustainability began to permeate the nation's collective consciousness (Hennig et al., 2020). The United Nations Conference on Environment and Development, held in Rio de Janeiro in 1992, served as a catalyst in propagating the notion of sustainability, with Germany emerging as one of the vanguards to embrace the concept wholeheartedly. Since that pivotal juncture, Germany has consistently implemented a panoply of policies and initiatives, meticulously designed to champion sustainable living (Hoffmann & Sprengel, 2010). Among these, the advent of the Renewable Energy Sources Act in 2000 holds prominence, as it established a robust legal framework for the development of renewable energy sources. This landmark policy proved instrumental in fomenting the utilization of wind and solar energy, playing a pivotal role in weaning Germany off fossil fuels. Another pivotal policy milestone was the introduction of the Energy Conservation Act in 2013, with a focused aim to augment energy efficiency in both buildings and industrial processes (Gölgeci & Kuivalainen, 2021). This innovative policy succeeded in curtailing Germany's energy consumption and greenhouse gas emissions, significantly contributing to the nation's proactive stance against climate change. Germany's prominence as a pioneer of sustainable living extends to the realm of transportation. The country has implemented an array of policies and initiatives with the purpose of promoting sustainable modes of travel. This entails subsidies for electric and hybrid vehicles and the meticulous expansion of an extensive network of bicycle paths and pedestrian-friendly infrastructure (Ivanov et al., 2021). Additionally, Germany has remained resolute in its commitment to fostering sustainable practices in agriculture, waste management, and urban development. The nation has fervently encouraged the adoption of organic farming practices and has instituted a comprehensive suite of policies to reduce waste and stimulate recycling. In urban areas, Germany's visionary approach to sustainable urban development prominently features public transportation prioritization, the proliferation of cycling infrastructure, and the nurturing of verdant communal spaces. In summation, Germany's remarkable trajectory in environmentalism and sustainable living traces back to the environmentally conscious ethos of the 1970s. This enduring commitment has been realized through an array of judiciously designed policies and initiatives, exemplified by the introduction of the Renewable Energy Sources Act, the Energy Conservation Act, and a suite of policies fostering sustainable transportation, agriculture, waste management, and urban development. These far-reaching efforts have not only diminished Germany's environmental footprint but have also established the nation as a trailblazer in the realm of environmental policy (Kaur, 2021).

**2.2 Germany's Bold Journey Toward Sustainable Energy**

Germany's relentless pursuit of sustainable energy solutions has earned it global recognition. The nation's ambitious objective of achieving 80% renewable energy by 2050 stands as a testament to its unwavering commitment to environmental sustainability (Kibert, 2021). In recent years, Germany has made substantial strides towards this goal, ushering in an era of renewable energy transformation. This transformative journey commenced in the early 2000s with the inception of the Renewable Energy Sources Act, a groundbreaking policy that incentivized the development of wind and solar energy sources. Since then, Germany has emerged as a frontrunner on the world stage, harnessing a significant portion of its energy from renewable sources, including wind, solar, and biomass. However, as with any paradigm shift, this transition to renewable energy has ushered in both challenges and opportunities. One of the paramount challenges faced by Germany is the intermittent nature of renewable energy sources, which are susceptible to the vagaries of weather patterns and various external factors. To surmount this challenge, Germany has directed its efforts toward energy storage solutions, such as advanced battery storage and pumped hydro storage. These technologies enable the storage of surplus energy during periods of abundance and its release when the demand necessitates it, mitigating the intermittent issue. Another challenge revolves around the impact of renewable energy integration on the country's energy grid. The influx of renewable sources into the grid has rendered it increasingly complex, necessitating more sophisticated management systems. In response, a strategic investment in smart grid technology, which serves to streamline energy flows and balance the supply-demand equation, ensuring grid reliability and efficiency in needed (Nawaser, 2015). Amid these challenges, Germany has uncovered a wealth of opportunities through its transition to renewable energy. Foremost among them is the potential to stimulate economic growth and job creation, particularly in rural areas blessed with abundant wind and solar resources. Additionally, this paradigm shift diminishes the nation's reliance on fossil fuels, bolstering energy security and reducing greenhouse gas emissions. Germany's steadfast commitment to renewable energy is a cornerstone of its broader sustainability agenda. While it has not been without its difficulties, including the intermittent nature of renewable sources and grid complexity, this transition presents substantial opportunities for economic growth, job creation, and enhanced energy security. Germany's unyielding journey towards sustainable energy is a shining example of proactive environmental stewardship and forward-thinking policy innovation (Lieder & Rashid, 2021).

**2.3 Elevating Sustainable Energy in Germany: Challenges and Opportunities**

Germany's resolute pursuit of transitioning to renewable energy sources has garnered international acclaim. The nation's audacious target of achieving 80% renewable energy by 2050 is emblematic of its unwavering commitment to sustainability, and in recent years, remarkable headway has been made in this regard (Masi, 2017). The journey towards renewable energy adoption in Germany commenced in the early 2000s, catalyzed by the groundbreaking Renewable Energy Sources Act. This pioneering legislation incentivized the development of wind and solar energy sources, laying the foundation for Germany to become a global exemplar in the field of renewable energy. Today, a substantial portion of the country's energy derives from wind, solar, and biomass sources, underscoring the nation's transformative transition. Nevertheless, this transition has not been without its share of challenges and opportunities. A central challenge pertains to the intermittent of renewable energy sources, rendering them susceptible to the capriciousness of weather patterns and other external factors. In response to this challenge, Germany has channeled its efforts into energy storage solutions, encompassing advanced battery storage and pumped hydro storage (Fiksel & Croxton, 2021). These innovations facilitate the accumulation of surplus energy during periods of abundance and its strategic release when demand is necessary, effectively addressing the intermittent issue. As Daneshmandi et al., (2023) and Drach-Zahavy & et al. (2004) discuss, In recent years, it has become common among researchers and scholars that innovation is crucial for organizations. Additionally, the integration of renewable energy sources into the nation's energy grid has presented a distinctive challenge. The augmentation of renewable sources has intensified the grid's complexity, necessitating the implementation of more sophisticated management systems. Germany has risen to this challenge by investing in smart grid technology, which adeptly manages energy flows, fine-tunes supply-demand dynamics, and ensures the grid's reliability and efficiency (Chen et al., 2021; Jahanshahi et al., 2019). Despite these challenges, Germany's transition to renewable energy has ushered in a trove of opportunities. Foremost among them is the potential for job creation and economic growth, particularly in rural regions blessed with abundant wind and solar energy resources. Furthermore, the shift away from fossil fuels enhances the nation's energy security, while concurrently reducing greenhouse gas emissions. In summary, Germany's unwavering commitment to renewable energy forms an indispensable pillar of its overarching sustainability agenda. While fraught with challenges such as the intermittent of renewable energy sources and grid complexity, this transition has also presented significant opportunities, including economic growth, job creation, and enhanced energy security. Germany's path to sustainable energy stands as a compelling model for other nations striving to reduce their carbon footprint and enhance their energy security (Pires et al., 2011).

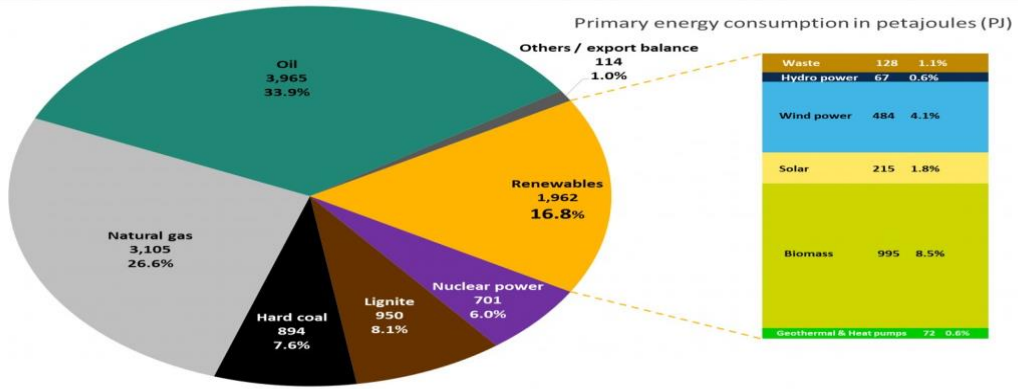

**Fig. 1.** German energy mix.

## 2.4 Nurturing Sustainable Transportation in Germany: A Complex Landscape

Germany's transportation sector bears a substantial responsibility for the nation's greenhouse gas emissions, with cars and trucks emerging as the predominant contributors. Consequently, the promotion of sustainable transportation stands as an indispensable facet of Germany's multifaceted approach to mitigating its environmental impact (Shishodia et al., 2022). In pursuit of this vital goal, Germany has orchestrated a medley of policies and initiatives, each designed to foster eco-conscious modes of mobility. Among the most pivotal of these initiatives is the resolute endorsement of public transportation. Germany boasts a well-developed public transportation network, encompassing buses, trains, and trams, that presents a convenient and cost-effective alternative to private vehicle usage. This comprehensive system actively encourages people to forego personal cars, thereby reducing emissions and mitigating the environmental impact of transportation. Simultaneously, Germany has embarked on a dedicated journey to enhance the cycling infrastructure. The country has made notable investments in the development of bike lanes and bike parking facilities, designed to stimulate cycling as a sustainable mode of transportation (Gölgeci & Kuivalainen, 2019). These investments not only reduce the reliance on cars but also promote a healthier, more active lifestyle. Another significant policy focus centers on the promotion of electric vehicles (EVs). Germany has set an ambitious target of having one million EVs on the road by 2020, underpinned by a wide array of incentives intended to expedite their adoption. These incentives encompass tax breaks for EV owners, subsidies for EV purchases, and substantial funding for the establishment of EV charging infrastructure. This holistic approach aims to accelerate the transition to cleaner and more sustainable transportation. Furthermore, Germany has implemented policies to reduce the environmental impact of traditional cars and trucks, including emissions standards that cap the amount of pollution these vehicles can emit. Additionally, the country has introduced a toll system for trucks, an innovative approach that encourages the utilization of more fuel-efficient vehicles while diminishing emissions from heavy-duty transport. While these policies and initiatives have demonstrated efficacy in promoting sustainable transportation in Germany, the journey is far from over. A fundamental challenge lies in the persistent dominance of car culture, which can complicate the shift to more sustainable modes of transportation. Moreover, the transition to electric vehicles has progressed more slowly than anticipated, attributed partly to concerns regarding the cost and range of EVs. The promotion of sustainable transportation has assumed a pivotal role in Germany's holistic approach to minimizing its environmental impact. Through an amalgamation of policies, including the promotion of public transportation, the development of cycling infrastructure, and the endorsement of electric vehicles, Germany endeavors to usher in more sustainable modes of transportation. While confronting enduring challenges, these efforts have nonetheless proven effective in advancing the cause of sustainable transportation within the country (Zhao et al., 2022).

2.5 Nurturing Sustainable Agriculture in Germany: Tilling a Greener Path

Agriculture, while fundamental to human existence, has long been associated with environmental degradation and greenhouse gas emissions (Bockreis & Steinberg, 2015). In Germany, the agricultural sector accounts for roughly 8% of the nation's greenhouse gas emissions and significantly impacts water and soil quality. To combat these challenges, Germany has instituted a comprehensive array of policies and initiatives designed to promote sustainable agriculture and reduce its environmental footprint. At the heart of Germany's sustainable agriculture drive lies a robust endorsement of organic farming. Organic farming diverges from conventional practices by eschewing synthetic pesticides and fertilizers and, instead, places a premium on soil health and biodiversity. To bolster this approach, Germany extends support to organic farming through a suite of subsidies and incentives. These incentives encompass funding for research and development, as well as direct subsidies for organic farmers. Organic farming has not only reduced the use of harmful chemicals but also fostered healthier soils and greater biodiversity. Another pivotal policy thrust is the promotion of sustainable land use. Germany has implemented a series of policies aimed at safeguarding and rehabilitating ecosystems. Wetlands and forests, crucial to the nation's environmental balance, are under protection. Additionally, Germany has championed sustainable land use practices such as crop rotation and conservation tillage, techniques that mitigate soil erosion and enhance soil quality. To address the environmental impact of livestock farming, Germany has adopted policies designed to encourage sustainable animal husbandry practices. Stringent animal welfare standards ensure humane treatment, preventing cruel or inhumane conditions for animals. Furthermore, Germany's focus on sustainable feed production promotes the use of locally sourced feed and explores alternative sources like insects, thereby reducing the ecological footprint of livestock farming. However, challenges persist in the pursuit of sustainable agriculture. High meat demand, which can amplify the environmental toll of livestock farming, continues to be a pressing issue. Additionally, the use of synthetic fertilizers in conventional farming remains a significant source of environmental pollution, necessitating further efforts to promote sustainable soil management practices (Zhao et al., 2022).

**3. Conclusion**

This paper has reviewed the literature on the role of sustainability in Germany, focusing on key sectors such as energy, transportation, agriculture, waste management, and urban development. Germany has implemented various policies and initiatives to advance sustainability across these domains, driven by its commitment to reduce greenhouse gas emissions, enhance energy efficiency, and protect the environment. However, the paper has also identified some ongoing challenges and areas for improvement, such as reducing emissions from the transportation sector, expanding sustainable farming practices, and reducing waste at its source. It is suggested that Germany can broaden its focus on sustainability to other industries, such as construction and manufacturing, and adopt sustainable practices and technologies to reduce their environmental impact. Moreover, Germany's dedication to sustainability is commendable and serves as a model for other nations. The country's notable achievements in transitioning to renewable energy, promoting sustainable transportation, fostering eco-friendly agriculture, and advancing waste management principles illustrate its commitment to a more sustainable future. While challenges remain, Germany's proactive stance in addressing these issues offers hope for a greener and more sustainable world.